\documentclass[twoside]{mhd_clean}

\usepackage[utf8]{inputenc}
\usepackage[english]{babel}
\usepackage{amsmath}
\usepackage{amssymb}
\usepackage{graphicx,color}

\usepackage{lineno}
\modulolinenumbers[5]

\title{A spectral/hp element MHD solver}

\author{Alexander~V. Proskurin\inst{1},Anatoly~M. Sagalakov\inst{2}}

\institute{Altai State Technical University, 656038, Russian Federation, Barnaul, Lenin prospect,46, k210@list.ru
\and
Altai State University, 656049, Russian Federation, Barnaul, Lenin prospect, 61 }

\begin{document}

\maketitle

\begin{abstract}
A new MHD solver, based on the \texttt{Nektar++} spectral/hp element framework, is presented in this paper. The velocity and electric potential quasi-static MHD model is used. The Hartmann flow in plane channel and its stability, the Hartmann flow in rectangular duct, and the stability of Hunt's flow are explored as examples.  Exponential convergence is achieved and the resulting numerical values were found to have an accuracy up to $10^{-12}$ for the state flows compared to an exact solution, and $10^{-5}$ for the stability eigenvalues compared to independent numerical results.
\end{abstract}

\sloppy

\section{Introduction}

It is well known that high-order methods have good computational properties, fast convergence, small errors, and the most compact data representation. For many problems in hydrodynamics, high-order methods are necessary. Such problems include the time-dependent simulation of transient flow regimes and the investigation of hydrodynamic stability. Of course, turbulent flows can be investigated using low accuracy schemes (achieving low precision results), but in most cases problems in channels of hydrodynamic stability  require the use of spectral methods. The classical Orr-Sommerfield equation has a small parameter $\frac{1}{Re}$ at the highest derivative which causes rapidly oscillating solutions. The first numerical calculation of eigenvalues for this equation \cite{thomas:1953} used a high-order finite difference scheme. Later, Orszag \cite{orszag:1971}  achieved more accurate results and explained the convenience of using high-order methods for problems of flow stability. A recent review of flow stability in complex geometries and the advantages and disadvantages of high-order methods can be found in \cite{theofilis:2011:highorder, theofilis:2011:global}.

\section{Problem formulation}

Consider a flow of incompressible, electrically conducting liquid in the presence of an imposed magnetic field. We suppose that $Re_m \ll Re$. In this case a magnetic field generated by the fluid movement does not affect the imposed magnetic field. This is correct for most engineering applications \cite{lee2001magnetohydrodynamic}.  It is now possible to write the Navier-Stokes equation in the form
\begin{equation}
\label{article19.NS_Syst_MagForce}
\begin{aligned}
\frac{\partial \boldsymbol{v}}{\partial{t}}+\left( \boldsymbol{v} \nabla \right)\boldsymbol{v} &= -\frac{1}{\rho}\nabla p + \nu \Delta \boldsymbol{v} + \boldsymbol{F}(\boldsymbol{v},\boldsymbol{H}),\\
div \boldsymbol{v} &= 0,
\end{aligned}
\end{equation}
where $\boldsymbol{v}$ is the velocity, $p$ is the pressure, $\nu$ is the viscosity, $\rho$ is the density, $\boldsymbol{F}$ is the magnetic force, and $\boldsymbol{H}$ is the magnitude of the imposed magnetic field.

Ohm's law is:
\begin{equation}
\label{article19.j_eq}
\boldsymbol{j} = \sigma\left( -\nabla\varphi+\boldsymbol{v}\times\boldsymbol{H}  \right),
\end{equation}
where $\boldsymbol{j}$ is the density of electirc current, $\varphi$ is the electric potential, and $\sigma$ is the conductivity. Using the law of conservation of electric charge ($div \boldsymbol{j} = 0$), it is possible to derive the equation for electric potential as:
\begin{equation}
\label{article19.ElPot_eq}
\Delta \varphi = \nabla(\boldsymbol{v}\times\boldsymbol{H}).
\end{equation}

The system (\ref{article19.NS_Syst_MagForce}) can be written in the form: 
\begin{equation}
\label{article19.WeakMHD_Syst}
\begin{aligned}
\frac{\partial \boldsymbol{v}}{\partial{t}}+\left( \boldsymbol{v} \nabla \right)\boldsymbol{v} &= -\nabla p + \frac{1}{Re} \Delta \boldsymbol{v} + St \left( -\nabla\varphi+\boldsymbol{v}\times\boldsymbol{H}  \right)\times\boldsymbol{H} ,\\
&div \boldsymbol{v} = 0,\\
&\Delta \varphi = \nabla(\boldsymbol{v}\times\boldsymbol{H}),
\end{aligned}
\end{equation}
where $Re = \frac{L_0V_0}{\nu}$ is the Reynolds number, $St=\frac{\sigma H_0^2 L_0}{\rho V_0}$ is the magnetic interaction parameter (Stuart number), and $L_0$, $V_0$, and $H_0$ represent the scales of length, velocity and magnitude of the imposed magnetic field, respectively. The system (\ref{article19.WeakMHD_Syst}) is also known as the MHD system in quasi-static approximation in electric potential form. This system is widely used in theoretical investigations and accurately approximate many cases of liquid metal flows (for example, see the appropriate discussion and reference in \cite{krasnov2011comparative}).

The boundary condition for velocity at walls have the form:
\begin{equation}
\label{article19.BounCond_V}
\boldsymbol{v} = 0,
\end{equation}
and the boundary condition for the electric potential at perfectly conducting walls is:
\begin{equation}
\label{article19.BounCond_phiD}
\varphi = const.
\end{equation}
The boundary condition for insulating walls is:
\begin{equation}
\label{article19.BounCond_phiN}
\frac{\partial \varphi}{\partial \boldsymbol{n}} = 0.\\
\end{equation}

\section{Numerical technique overview}

Our new MHD solver has been developed on the basis of an open source spectral/hp element framework \texttt{Nektar++} \cite{cantwell:2015, karniadakis:2013}.
The incompressible Navier-Stokes solver (\texttt{IncNavierStokesSolver}) from the framework has been taken as the source for the MHD solver.

\texttt{IncNavierStokesSolver} uses the velocity correction scheme as described in \cite{karniadakis:2013,karniadakis:1991:high}, assuming the time grid $t_{0}, t_{1}, \ldots,t_{n-1},t_{n},t_{n+1}$. Using a first-order difference scheme, it is possible to define intermediate velocity $\tilde{\boldsymbol{v}}$ by the equation:
\begin{equation}
\label{article19.VelCorr_1st_inter_vel}
\frac{\tilde{\boldsymbol{v}}-\boldsymbol{v}_n}{\delta t} = -\left(\boldsymbol{v}_n \nabla \right)\boldsymbol{v}_n + \boldsymbol{F}_n+St \cdot \, \boldsymbol{v}_n\times \left( -\nabla\varphi_n+\boldsymbol{v}_n\times\boldsymbol{H}  \right),
\end{equation}
where $\boldsymbol{F}$ is a force acting of a fluid. At this stage, the \texttt{Nektar++} software allows force to be imposed which acts on the flow. The electrical potential is found by solving the equation $\Delta \varphi_n = \nabla(\boldsymbol{v}_n\times\boldsymbol{H})$ and, after this, the magnetic force in (\ref{article19.VelCorr_1st_inter_vel}) is calculated.
We define the second intermediate velocity as:
\begin{equation}
\label{article19.VelCorr_2nd_inter_vel}
\frac{\hat{\tilde{\boldsymbol{v}}}-\tilde{\boldsymbol{v}}}{\delta t} = - \nabla p_{n+1}.
\end{equation}
The Poisson equation
\begin{equation}
\label{article19.VelCorr_Poissin_eq}
\Delta p_{n+1} = \nabla \left(\frac{\tilde{\boldsymbol{v}}}{\delta t}\right)
\end{equation}
is immediately derived using $div \hat{\tilde{\boldsymbol{v}}} = 0 $. Thus, at this stage, the divergence-free condition is approximately satisfied. The boundary conditions for pressure are discussed in \cite{karniadakis:1991:high}. The last step of the velocity correction procedure is the equation:
\begin{equation}
\label{article19.VelCorr_n+1_vel}
\left(\Delta - \frac{Re}{\delta t}\right)\boldsymbol{v}_{n+1} = - \frac{Re}{\delta t} \tilde{\boldsymbol{v}} + Re\, \nabla p_{n+1},
\end{equation}
wich allows us to find the next time-step velocity $\boldsymbol{v}_{n+1}$. \texttt{Nektar++} can use first, second and third order schemes.

\section{The Hartmann flow in a plane channel}\label{article19.hartmannflowsection}

\begin{figure}[h]
\begin{center}
\includegraphics{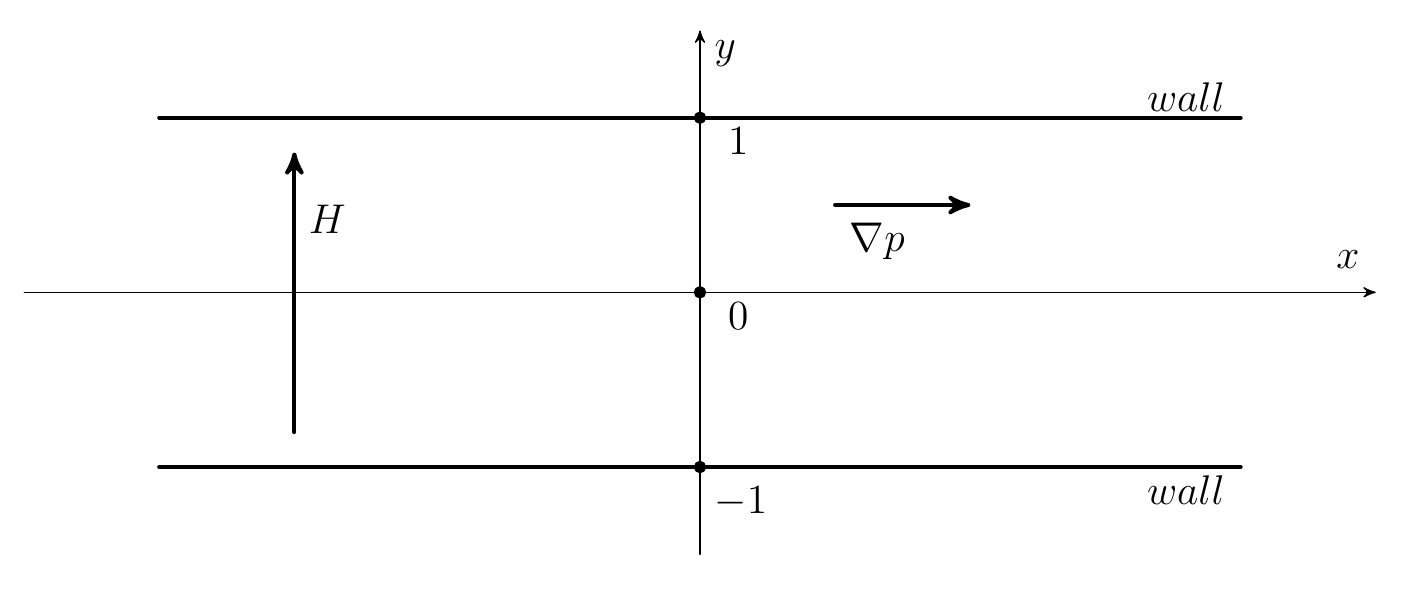} 
\end{center}
\caption{The plane channel}\label{article19.PlaneChannelPict}
\end{figure}

\begin{figure}[h]
\begin{center}
\includegraphics{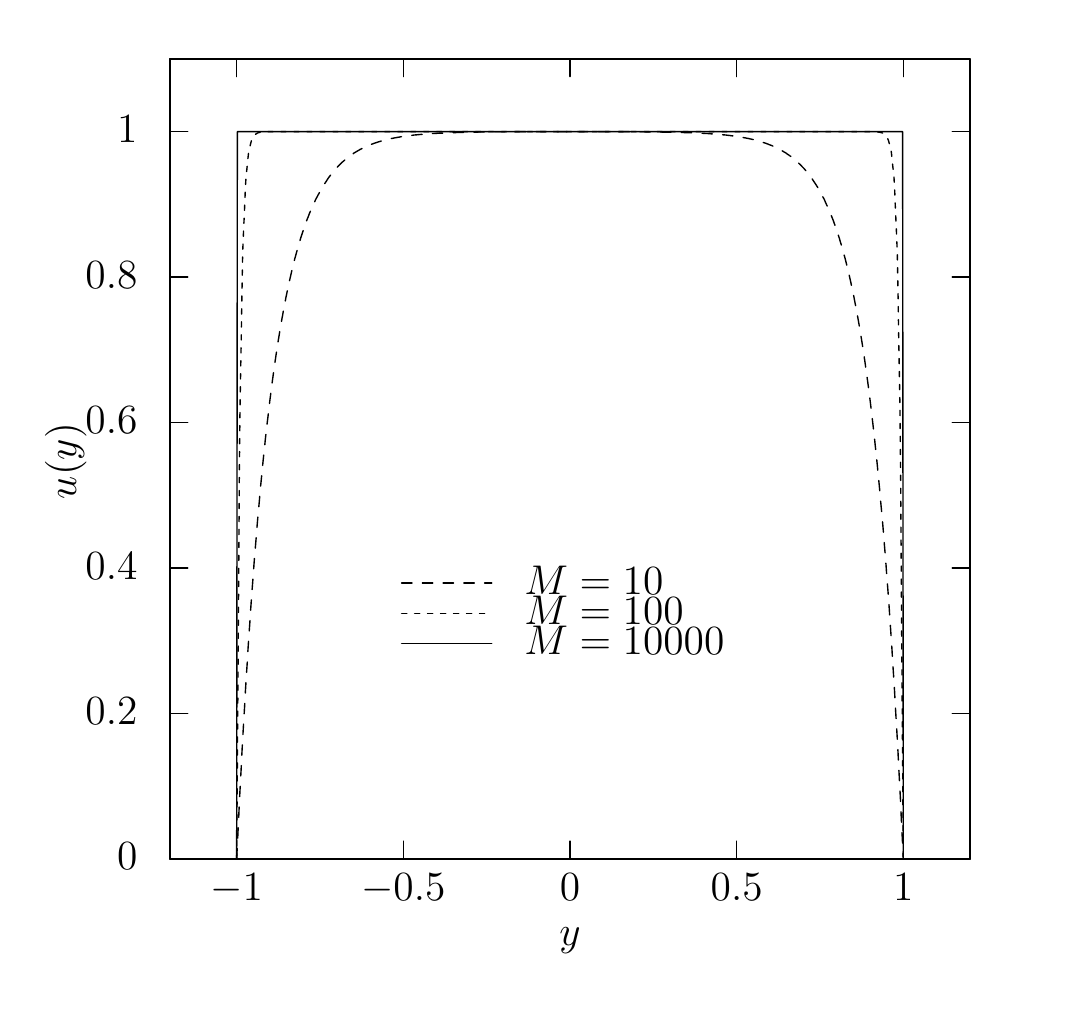} 
\end{center}
\caption{Velocity of the Hartmann flow at different values of M}\label{article19.VelocityHartmannChannelPict}
\end{figure}

Figure \ref{article19.PlaneChannelPict} illustrates a flow in a plane channel. Two parallel infinite planes are installed at points $y = \pm 1$. The liquid between the planes flows under a constant pressure gradient $\nabla p$ in a $x$ direction. The magnetic field $\boldsymbol{H}$ is perpendicular to the planes. This is the Hartmann flow in the plane channel. According to \cite{davidson2016introduction} the solution of (\ref{article19.WeakMHD_Syst}) is 
\begin{equation}
\label{article19.hartman_flow_1}
\frac{u(y)}{u(0)}=\frac{\cosh{\left (M \right )} - \cosh{\left (M y \right )}}{\cosh{\left (M \right )} - 1}
\end{equation}
where $M = \sqrt{St \cdot Re}$, $u(0)$ is a centreline velocity. Velocity graphs are shown in Figure \ref{article19.VelocityHartmannChannelPict} for $M=10,100,10000$.

\begin{figure}[h]
\parbox[c][4cm][c]{\textwidth}{\includegraphics[width=14cm,height=9cm]{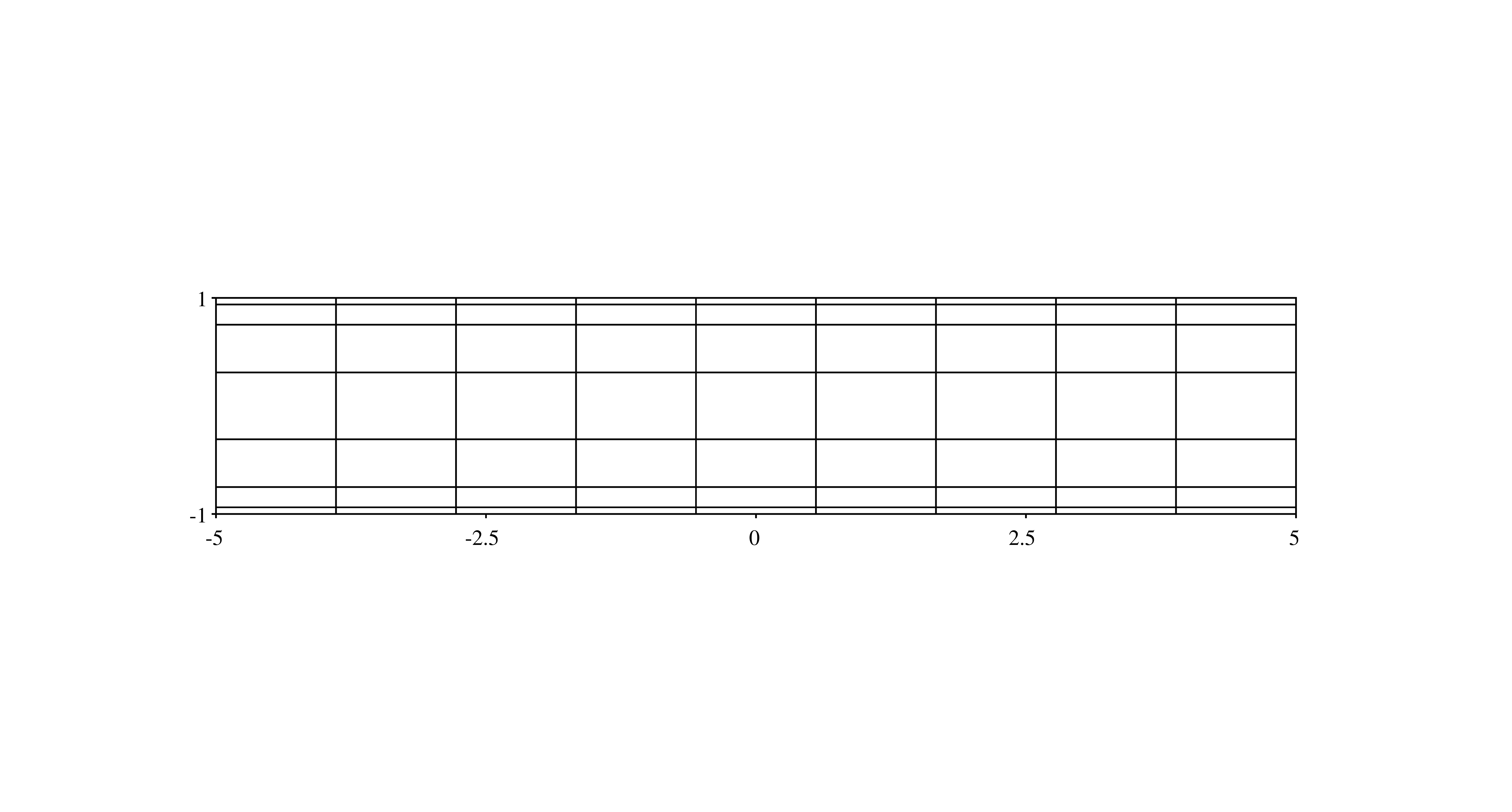} \rule{10cm}{0cm}}
\caption{A mesh for calculations of the Hartmann flow, $\beta=0.05$}\label{article19.HartmannChannelMeshPict}
\end{figure}

Figure \ref{article19.HartmannChannelMeshPict} shows a 2D mesh for numeric calculations of the flow. In Figure \ref{article19.VelocityHartmannChannelPict} one can see large gradients of velocity near the walls in cases of large $M$ and we should take special attention to this part of the flow. It is possible to increase accuracy by mesh concentration near the walls where there are large velocity gradients. This is the $h$-type solution refinement. The high-order method can increase accuracy by increasing the polynomials' order of an approximation, this is $p$-refinement. For the flow calculations we will combine both methods by using the order of approximation $p$ from $5$ to $25$ and mesh condensation near the walls with a coefficient $\beta$ ($\beta=1$ for a uniform grid).  

When it is supposed that the flow is two-dimensional, all functions are independent from $z$-coordinate and $v_z=0$. This proposition leads to the equation $\Delta \varphi = 0$,  this means that boundary conditions for $\varphi$ are not required.
The boundary conditions for velocity and pressure are  
\begin{equation}
\label{article19.BoudaryCondCalc}
\begin{aligned}
\boldsymbol{v} &= 0, \, \frac{\partial p}{\partial n} = 0 \, \text{at walls},\\
\frac{\partial \boldsymbol{v}}{\partial n}&= 0,\, \frac{\partial p}{\partial n} = 0 \, \text{at inflow and outflow}.\\
\end{aligned}
\end{equation}

In Table \ref{article19.HartmannChannelFlowResultConv_p}, maximum deviations from the exact solution are presented at $M=10\sim 10^4$ for different orders of polynomial approximation $p$. The state flow (\ref{article19.hartman_flow_1}) is calculated as a time-dependent flow with zero initial conditions. Table \ref{article19.HartmannChannelFlowResultConv_p} includes the running time of the solver on an AMD Ryzen Threadripper 1920X machine with 12 threads. 

\section{Flow in rectangular duct}
\begin{figure}[h]
\begin{center}
\includegraphics{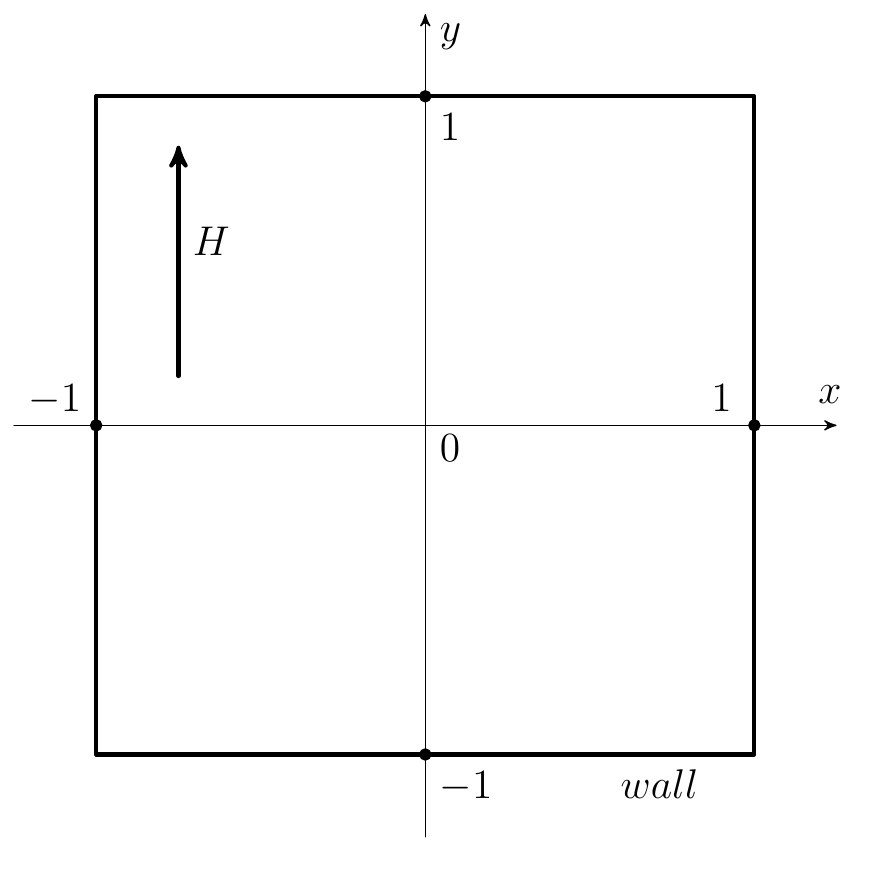}
\caption{The rectangular duct}\label{article19.RectangularDuct}
\end{center}
\end{figure}

Consider a steady flow in a rectangular duct. A sketch of this flow is shown in Figure \ref{article19.RectangularDuct}. The rectangle in Figure \ref{article19.RectangularDuct} is the cross section of the channel and a uniform magnetic field is applied vertically. Liquid flows under a constant pressure gradient, perpendicular to the plane of the diagram. This flow was investigated analytically in \cite{chang1961duct, shercliff1953steady}.

\begin{figure}[h]
\parbox[c][10cm][c]{\textwidth}{\includegraphics[width=14cm,height=7.5cm]{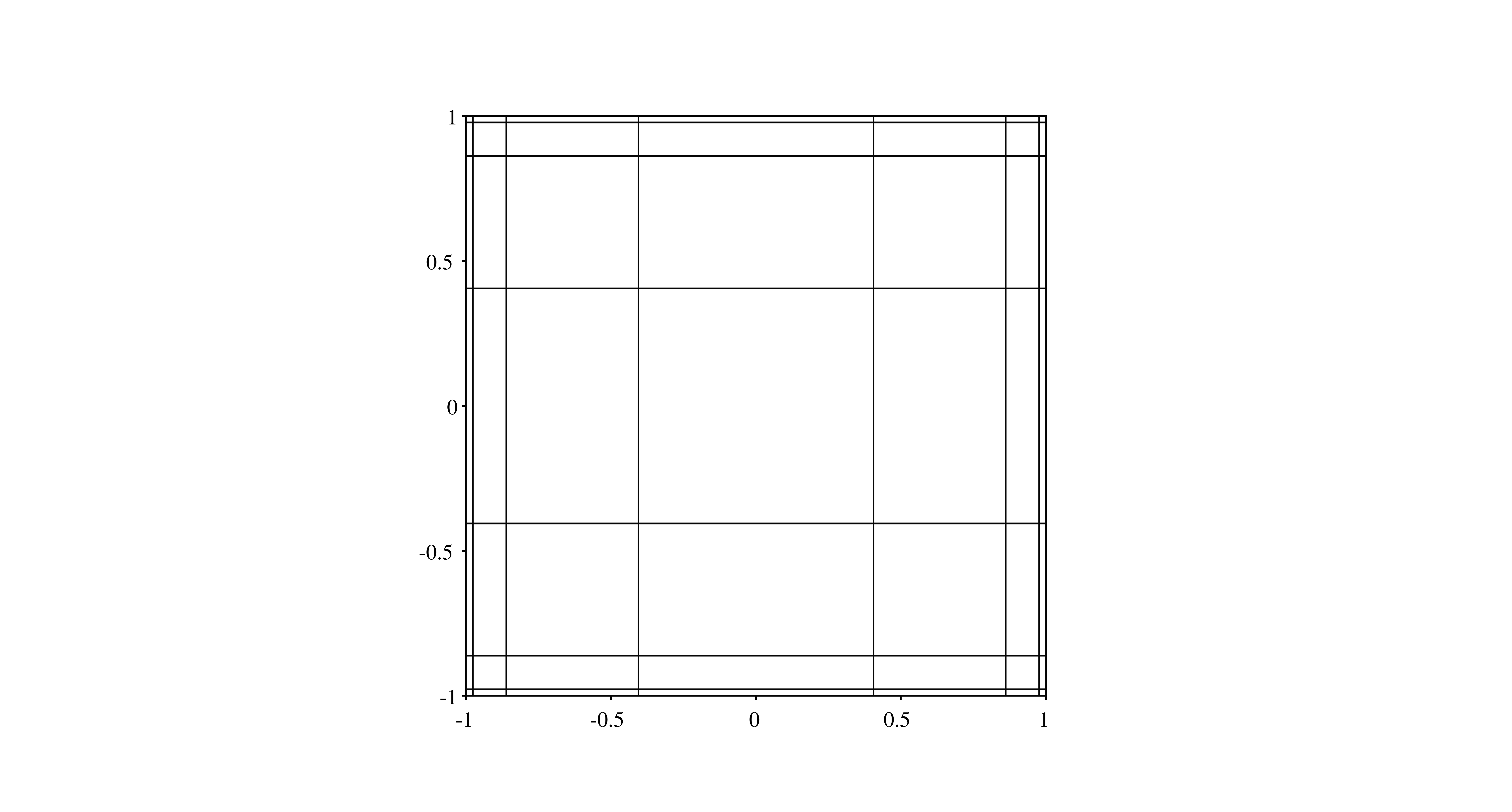}}
\caption{The mesh for calculations of the Hartmann flow in duct, $\beta=0.01$}\label{article19.HartmannDuctMeshPict}
\end{figure}

For flow computations in the rectangular duct, the authors use a mesh shown in Figure \ref{article19.HartmannDuctMeshPict}. \texttt{Nektar++} software allows us to set up $3D$ problems where the homogeneous direction is $z$, the number of Fourier modes is $2$(the minimal possible value).

\begin{figure}[h]
\begin{center}
\includegraphics{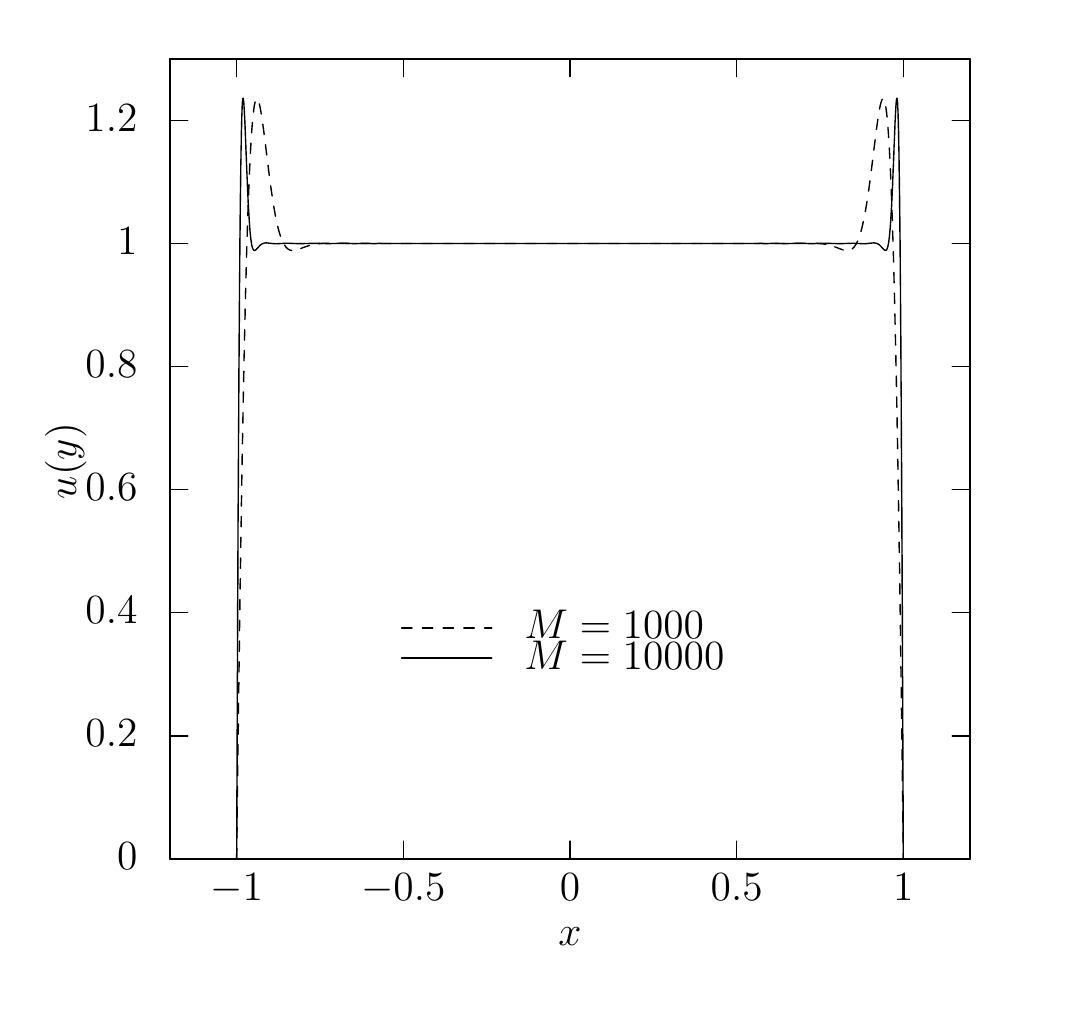}
\caption{Velocity of the Hartmann flow in rectangular duct at line $y=0$ for case of ideal electro conducting walls,  $M=10^3,\,10^4$}\label{article19.DuctProfile}
\end{center}
\end{figure}

The velocity convergence at points $(0.95,0.0)$ for $M=10^3$ and $(0.98,0.0)$ for $M=10^4$ is presented in Table \ref{article19.HartmannDuctFlowConvergence} for the case of perfectly electro-conducting walls. The table includes the mesh concentration coefficient $\beta$ and velocity values for different $p$ from $5$ to $25$, the time of calculation and memory usage, for an AMD Ryzen Threadripper 1920X machine with 12 threads. The velocity graph for $M=10^3$, $M=10^4$ is shown in Figure \ref{article19.DuctProfile}.

\section{The stability problem}
For a stability analysis let us decompose velocity, pressure and electric potential to form
\begin{equation}
\label{article19.linear_form}
\begin{aligned}
\boldsymbol{v} &= \boldsymbol{U}+\boldsymbol{v},\\
\varphi &= \varphi_0+\varphi,\\
p &= p_0 +p,
\end{aligned}
\end{equation}
where $U$, $\varphi_0$, and $p_0$ are values of a steady-state flow and $\boldsymbol{v}$, $\varphi$ and $p$ are small disturbances. The system (\ref{article19.WeakMHD_Syst}) becomes linear:
\begin{equation}
\label{article19.linearNS}
\begin{aligned}
\frac{\partial \boldsymbol{v}}{\partial{t}}+\left( \boldsymbol{U} \nabla \right)\boldsymbol{v}+\left( \boldsymbol{v} \nabla \right)\boldsymbol{U} &= -\nabla p + \frac{1}{Re} \Delta \boldsymbol{v} + St \left( -\nabla\varphi+\boldsymbol{v}\times\boldsymbol{H}  \right)\times\boldsymbol{H} ,\\
&div \boldsymbol{v} = 0,\\
&\Delta \varphi = \nabla(\boldsymbol{v}\times\boldsymbol{H}).
\end{aligned}
\end{equation}

From equations (\ref{article19.linearNS}) a linear operator $\boldsymbol{A}$ can be constructed:
\begin{equation}
\label{article19.stab_eigproblem}
\boldsymbol{v}(x,y,z,T) = \boldsymbol{A}(T)\boldsymbol{v}(x,y,z,0) = \lambda(T) \boldsymbol{v}(x,y,z,0),
\end{equation}
where $T$ is a time interval.
The linear operator $\boldsymbol{A}(T)$ is constructed numerically by the splitting procedure in the same way as in the case of  nonlinear equations (\ref{article19.WeakMHD_Syst}). In order to find an eigenvalue $\lambda(T)$, it is convenient to construct a Krylov subspace
\begin{equation}
\label{article19.stab_krylovsbspace}
K_n(\boldsymbol{A},\boldsymbol{v}_0) = span\{\boldsymbol{v}_0, \boldsymbol{A}(T) \boldsymbol{v}_0,  \boldsymbol{A}(T)^2 \boldsymbol{v}_0, \ldots,  \boldsymbol{A}(T)^{n-1} \boldsymbol{v}_0 \},
\end{equation}
where $\boldsymbol{A}(T)^i v_0$ is obtained by direct calculation $\boldsymbol{v}_1 = \boldsymbol{A}(T)\boldsymbol{v}_0$, $\boldsymbol{v}_2 = \boldsymbol{A}(T)\boldsymbol{v}_1$, $\ldots$. Further eigenvalue calculations are carried out by standard numerical algebraic techniques, such as the Arnoldi method. The eigenvalues are obtained by \texttt{Nektar++} in the form:
\begin{equation}
\label{article19.eigenval_form}
\lambda(T) = m \cdot e^{\theta i},\\
\end{equation}
and if $m>1$ then the flow is unstable. The time-independent growth is $\sigma = \frac{ln(m)}{T}$ and the time-independent frequency is $\omega=\frac{\theta}{T}$.

\section{Stability of the Hartmann Flow}
In Section \ref{article19.hartmannflowsection}, the Hartmann flow was considered. Now, we will explore the stability of this flow. We take a Hartmann 2D flow disturbance (\ref{article19.linear_form}) in the form:
\begin{equation}
\label{article19.stab_wavesol}
\left \{\boldsymbol{v},\varphi,p \right \} = \boldsymbol{q}(x,y) e^{i\alpha(x-C t)},
\end{equation}
where $\boldsymbol{q}(x,y)$  is the amplitude of disturbance, $\alpha=\frac{2\pi}{\gamma}$ is the wave-vector, $\gamma$ is the wavelength, $C=X+iY$ is the phase velocity of disturbance, $\alpha X=\omega$ is the frequency, $\alpha Y = \sigma$ is the growth of disturbance. When $\sigma \leq 0$, it means that the flow is stable. 

The disturbance form (\ref{article19.stab_wavesol}) is widely used in hydrodynamics stability analysis and leads to the eigenvalue problem, equivalent to (\ref{article19.stab_eigproblem}). As reference data, we take Takashima critical values \cite{takashima1996stability} for 2D disturbances.  
In Table  \ref{article19.Harmann_stab_table}, growths and frequencies of the disturbances are given for several cases. The values of $Re$ and $M$ 
are taken from the article \cite{takashima1996stability}, and these are critical values of Hartmann flow. The computational grid is shown in Figure \ref{article19.HartmannChannelMeshPict}; $nx$ and $ny$ are the number of cells in the horizontal and vertical directions. The length $L$ of the grid is set up by using a critical wave-vector  $\alpha_c = \frac{2\pi}{L}$. Boundary conditions at inlet and outlet are periodical. In Table \ref{article19.Harmann_stab_table} complete coincidence is observed with the reference data from \cite{takashima1996stability} and convergence of the eigenvalues is achieved up to $10^{-7}$. Additionally, the table includes the time and memory usage (for the AMD Phenom FX-8150 processor with 8 threads).

\section{Stability of Hunt's flow}
\begin{figure}[h]
\parbox[c][10cm][c]{\textwidth}{\includegraphics[width=14cm,height=7.5cm]{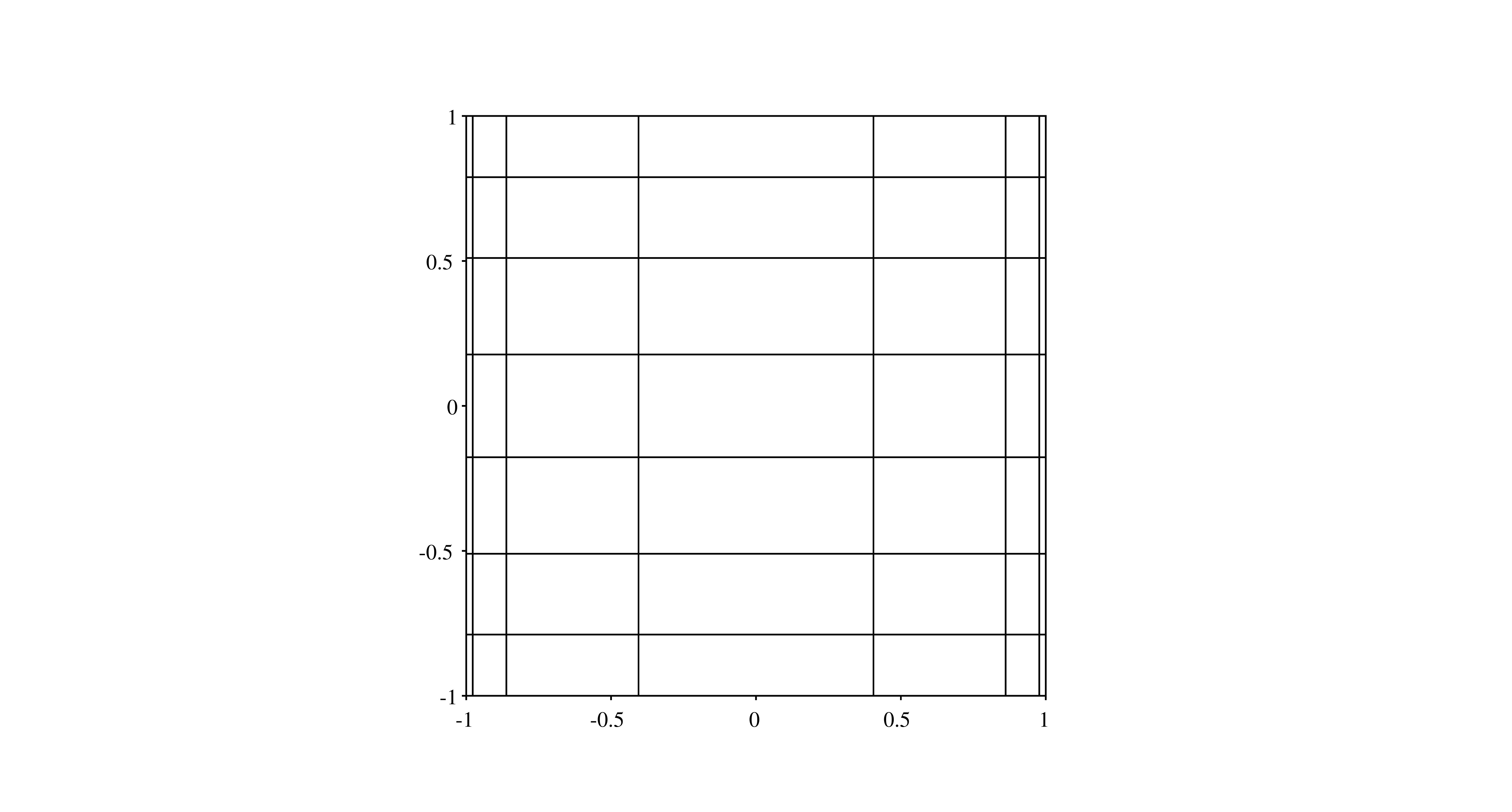}}
\caption{The mesh for calculations of Hunt's flow, $\beta_y=0.5$, $\beta_x=0.01$}\label{article19.HuntDuctMeshPict}
\end{figure}

\begin{figure}[h]
\begin{center}
\includegraphics[width=10cm]{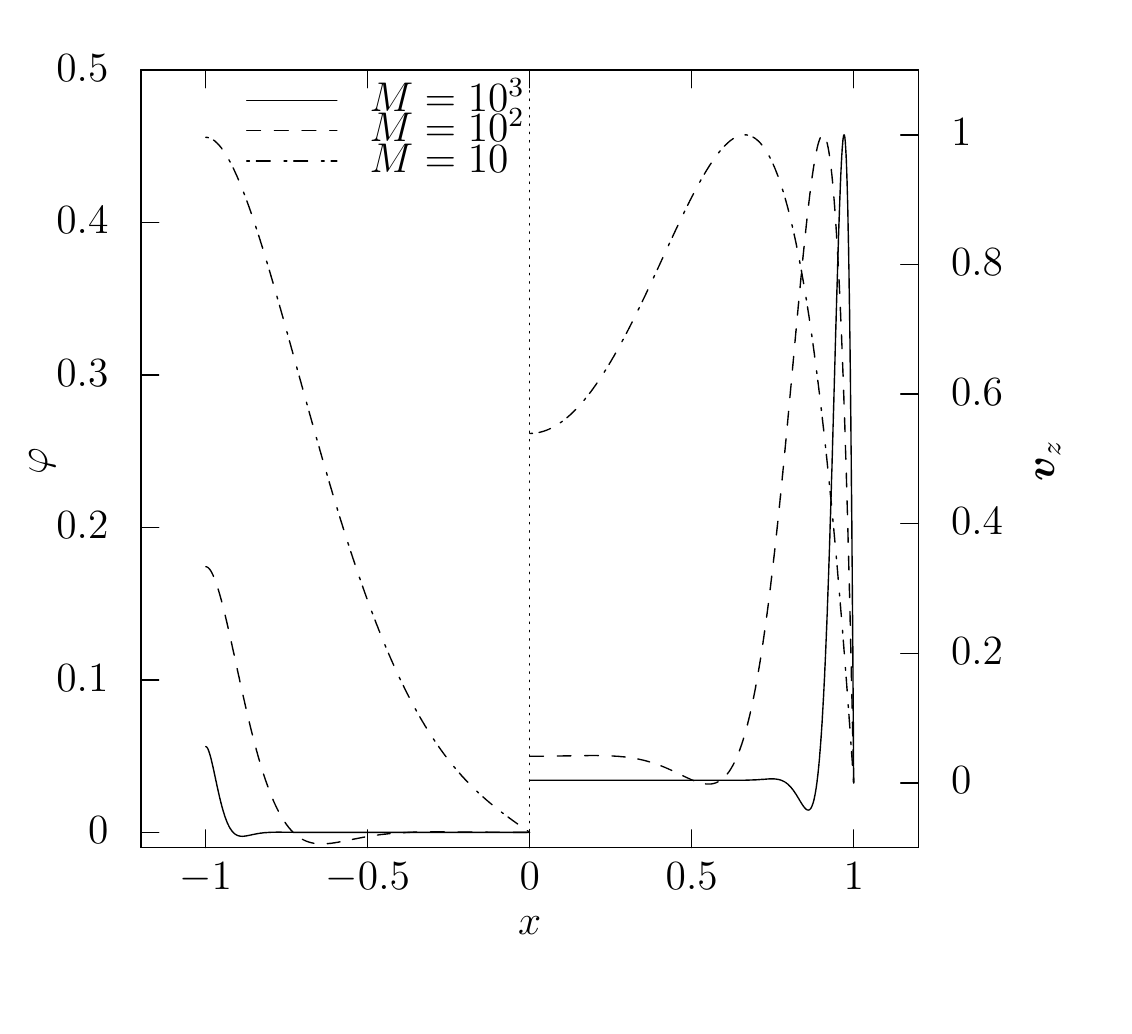}
\caption{Velocity $\boldsymbol{v}_z$ and electric potential $\varphi$ of Hunt's flow at line $y=0$,  $M=10,\,10^2,\,10^3$}\label{article19.HuntDuctProfile}
\end{center}
\end{figure}

Consider a steady flow in the rectangular duct (Figure \ref{article19.RectangularDuct}), where horizontal walls are perfectly electrically conducting and vertical walls are perfectly electrically insulating. The flow was investigated in \cite{hunt1965magnetohydrodynamic} and it is known as the Hunt's flow. A mesh for base flow and stability calculations is shown at Figure \ref{article19.HuntDuctMeshPict}. Figure \ref{article19.HuntDuctProfile} presents a graph of velocity over a line $y=0$ at $M=10,\,10^2\,10^3$. In Table \ref{article19.HantDuctFlowStability} our calculated eigenvalues are compared with reference values from article \cite{priede2010linear}; the time of calculation and the memory usage are presented. It is possible to see numerical convergence by increasing the order $p$ and match with the reference values from \cite{priede2010linear} up to $10^{-5}$, excluding the case $M=10^3$.

\section{Conclusion}
This article presents the spectral/hp element solver for MHD problems based on the \texttt{Nektar++} framework. The solver also makes it possible to investigate the stability of such flows. In order to demonstrate the solver's capacity, several examples were considered: the Hartmann flow in a plane channel and its stability, the Hartmann flow in a rectangular duct, and the stability of 
Hunt's flow. For the flows, it is easy to find steady-state solutions analytically, and these results were used as the reference test solutions. It was found that the margin of error decreases exponentially with an increasing degree of approximating polynomials, an accuracy $10^{-12}$ can be achieved. To estimate the costs of computer time and memory, these data were listed in the tables for several cases. The computational costs for the stability calculations are large. The first reason for this is the fact that a non-stationary algorithm was used, which allowed use of the non-stationary solver with small adaptations. To obtain eigenvalues with high accuracy, we should set a small time step. The second reason is that we are considering the test examples in 2D for the Hartmann flow and 3D for flows in duct. Usually, these problems can be reduced to the more simple cases described in \cite{takashima1996stability,priede2010linear,priede2012linear}, which can be investigated with much lower computational costs. In this article we demonstrated the accuracy of the method using the well investigated examples. 
In general, our numerical technique is intended for complex geometry flows where such simplifications are not possible.

\bibliographystyle{mhd}
\bibliography{reference1}

\begin{table}[p]
\caption{Maximum deviation from the exact solution (\ref{article19.hartman_flow_1}), $M=10\sim10^4$}\label{article19.HartmannChannelFlowResultConv_p}
\begin{center}
\small
\begin{tabular}{|c||l|l||l|l||l|l||l|l|}
\hline
&\multicolumn{2}{c||}{$\beta=0.5$}     &\multicolumn{2}{|c||}{$\beta=0.05$}     &\multicolumn{2}{|c||}{$\beta=0.005$}      &\multicolumn{2}{|c}{$\beta=0.0005$}\\
\hline
$p$     &$M=10$     &$t$,m:s    &$M=100$   &$t$,m:s&$M=10^3$   &$t$,m:s &$M=10^4$   &$t$,m:s  \\
\hline
5	&0.00035269	    &0:00.50	&0.01206	    &0:00.71	&0.0659132	    &0:00.48	&0.116924	    &0:01.29\\
10	&8.55003E-10	&0:02.93	&3.67734E-06	&0:04.51	&0.000550293	&0:02.78	&0.00731443	    &0:07.09\\
15	&6.91253E-13	&0:21.02	&1.50558E-08	&0:16.32	&7.01068E-07	&0:12.65	&0.000215073	&0:23.43\\
20	&7.76823E-13	&0:41.80	&4.32387E-11	&0:53.42	&9.98051E-10	&0:33.59	&4.2223E-05	    &0:58.08\\
25	&8.36775E-13	&1:39.52	&4.49063E-11	&1:58.71	&7.00766E-10	&1:15.66	&4.10316E-05	&3:36.74\\
\hline
\end{tabular}
\end{center}
\end{table}

\begin{table}[h]
\caption{Convergence of velocity values at point $(x,y)$, $M=10^3,\,10^4$}\label{article19.HartmannDuctFlowConvergence}
\begin{center}
\begin{tabular}{|c||l|r||l|r||r|}
\hline
&\multicolumn{2}{|c||}{$\beta=0.0005$, $(x,y)=(0.95,0.0)$}&  \multicolumn{2}{c||}{$\beta=0.0001$, $(x,y)=(0.98,0.0)$}&\\
\hline
$p$ & $M=10^3$ & $time$, m:s & $M=10^4$ & $time$, m:s & \parbox{1.1cm}{\tiny memory usage, Gb}\\
\hline
5	&1.197537688505880	&0:54.71	&1.105637945879100	&2:35.64	&0.041\\
10	&1.214652059488780	&1:31.85	&1.242371828471130	&9:29.99	&0.115\\
15	&1.212061289609800	&5:12.96	&1.234533071659520	&24:40.63	&0.358\\
20	&1.212043863156160	&9:30.71	&1.234851578015370	&1h 12:50	&0.962\\
25	&1.212044479951230	&21:53.03	&1.234834938798250	&2h 30:03	&2.189\\
\hline
\parbox{1cm}{\tiny reference values} & 1.21204510 & &1.2348750 & & \\
\hline
\end{tabular}
\end{center}
\end{table}

\begin{table}[h]
\caption{Eigenvalues of the Hartmann flow}\label{article19.Harmann_stab_table}
\begin{center}
\tiny
\begin{tabular}{|l|c|c|c|c|l|c|c|c|}
\hline
$Re$        &$M$ & $\alpha$  & $n_x \times n_y$ & p  & $\frac{\sigma}{\alpha}$ & $\frac{\omega}{\alpha}$ & time    & memory,GB\\
\hline
 10016.2621 & 1  & 0.971828  & $6\times 3 $     & 10 & 0.233652452491593       & 4.14E-04  & 1h18:12 & 0.07     \\
 -          & -  & -         & -                & 20 & 0.235517912202571       & 1.55E-06  & 3h28:46 & 0.33     \\

 -          & -  & -         & $8\times 6 $     & 10 & 0.235518158254341       & -7.67E-07 & 1h47:42 & 0.11     \\
 -          & -  & -         & -                & 15 & 0.235518939061233       & 8.97E-08  & 4h35:36 & 0.30     \\
 -          & -  & -         & -                & 20 & 0.235518994976477       & 6.38E-08  &11h03:43 & 0.77     \\
\hline
 -          & -  & -         &\multicolumn{2}{|c|}{reference value} &0.235519                 &\multicolumn{3}{|l|}{}\\
\hline
 28603.639  & 2  & 0.927773  & $8\times 6 $     & 10 & 0.192137721274493       & 1.66E-06  & 3h19:01 & 0.13     \\
\hline
 -          & -  & -         &\multicolumn{2}{|c|}{reference value} & 0.192133                &\multicolumn{3}{|l|}{}\\
\hline
 65155.21   & 3  & 0.958249  & $8\times 6 $     & 20 & 0.169030377438432       & 1.16E-07  & 15h41:21 & 0.77     \\
\hline
 -          & -  & -         &\multicolumn{2}{|c|}{from \cite{takashima1996stability}} & 0.169030                &\multicolumn{3}{|l|}{}\\
\hline

\end{tabular}
\end{center}
\end{table}


\begin{table}[h]
\caption{Eigenvalues of the Hunt flow, $M=10,\,10^2,\,10^3$}\label{article19.HantDuctFlowStability}
\begin{center}
\tiny
\begin{tabular}{|c||l|l|r||l|l|r||l|l|r||c|}
\hline
&\multicolumn{3}{|c||}{$M=10$,$\beta_x=0.05$, $\beta_y=0.1$}&\multicolumn{3}{|c||}{$M=10^2$,$\beta_x=0.01$, $\beta_y=0.5$}&\multicolumn{3}{|c||}{$M=10^3$,$\beta_x=0.01$, $\beta_y=0.5$}&\\
\hline
$p$ &$\frac{\sigma}{\alpha}$&$\frac{\omega}{\alpha}$&time &$\frac{\sigma}{\alpha}$&$\frac{\omega}{\alpha}$&time &$\frac{\sigma}{\alpha}$&$\frac{\omega}{\alpha}$&time&\parbox{0.5cm}{\tiny mem, Gb}\\
\hline
5	&0.7495498	&-0.5494e-2	& 7:02	    &0.8109660	&-0.134660	&5:48	    &0.4576533	&2.41323e-2	&2:12	&0.042\\
10	&0.7579881	&-0.3446e-3	& 15:13	&0.4912104	&-0.6873e-2	&42:59	    &0.4799186	&0.08964e-2	&14:12	&0.118\\
15	&0.7579733	&-0.3348e-3	& 1h27:24	&0.4907436	&-0.8033e-2	&1h50:50	&0.4784210	&0.14098e-2	&2h21:50	&0.365\\
20	&0.7579599	&-0.3311e-3	&5h57:42	&0.4907460	&-0.8030e-2	&2h42:15    &0.4784201	&0.14121e-2	&3h50:34	&0.972\\
\hline
\parbox{0.5cm}{from\\ \cite{priede2010linear}}	&0.7579413	&-0.3337e-3	&           &0.4907415	&-0.8028e-2	&	        &0.5053902	&0.14170e-2	&&\\	
\hline
\end{tabular}
\end{center}
\end{table}

\end{document}